\documentclass[%
 reprint,
 amsmath,amssymb,
 aps,
]{revtex4-2}

\usepackage{graphicx}% Include figure files
\usepackage{dcolumn}% Align table columns on decimal point
\usepackage{bm}% bold math
\usepackage{braket}
\usepackage{changes}

\usepackage{comment}
\usepackage{hyperref}       % hyperlinks
\hypersetup{
    colorlinks=true,
    linkcolor=blue,
    filecolor=magenta,      
    urlcolor=cyan,
}
\usepackage{color,soul}

\begin{document}

\preprint{APS/123-QED}

\title{Dissipative structures in topological lattices of nonlinear optical resonators}

\author{Aleksandr K. Tusnin$^{1,2}$}
 \email{aleksandr.tusnin@epfl.ch}
\author{Alexey M. Tikan$^{1,2}$}%
 \email{alexey.tikan@epfl.ch}
\author{Kenichi Komagata$^{1,3}$}
\author{Tobias J. Kippenberg$^{1,2}$}%
 \email{tobias.kippenberg@epfl.ch}
\affiliation{%
 $^1$Institute of Physics, Swiss Federal Institute of Technology Lausanne (EPFL), Lausanne, Switzerland\\
 $^2$Center for Quantum Science and Engineering, EPFL, Lausanne, Switzerland \\
$^3$Present address: Laboratoire Temps-Fréquence, Avenue de Bellevaux 51, 2000 Neuchâtel, Switzerland
}%

\date{\today}% It is always \today, today,

\begin{abstract}
We theoretically study the dynamics and spatio-temporal pattern formation of driven lattices of nonlinear optical microresonators and analyze the formation of dissipative structures, in particular dissipative Kerr solitons.  We consider both equally coupled one-dimensional chains, as well as the topological Su-Schrieffer-Heeger model.  We show the complexity of the four-wave mixing pathways arising in these systems with the increasing dimensionality due to the combined spatial and synthetic
    frequency dimension of each resonator, and show that it can be modeled using a two-dimensional variant of the Lugiato-Lefever equation. We demonstrate the existence of two fundamentally different dynamical regimes in
    one-dimensional chains - elliptic and hyperbolic - inherent to the system.  In the elliptic regime, we generate hexagonal patterns and a two-dimensional dissipative Kerr soliton corresponding to the global spatio-temporal mode-locking and discuss its similarity to edge-state solitons in the two-dimensional Haldane topological lattice. We find that the presence of the second dimension leads to the observation of regularized wave collapse. Furthermore, we study similarities and differences
    between a one-dimensional topological lattice and a single cavity and analyze nonlinearly induced edge-to-bulk scattering in the Su-Schrieffer-Heeger model. 
Moreover, we show that soliton formation can both be impaired in trivial but, importantly, also topologically protected bands due to nonlinear bulk edge scattering. 
\end{abstract}
\maketitle

\begin{figure*}
    \centering
    \includegraphics[width=\textwidth]{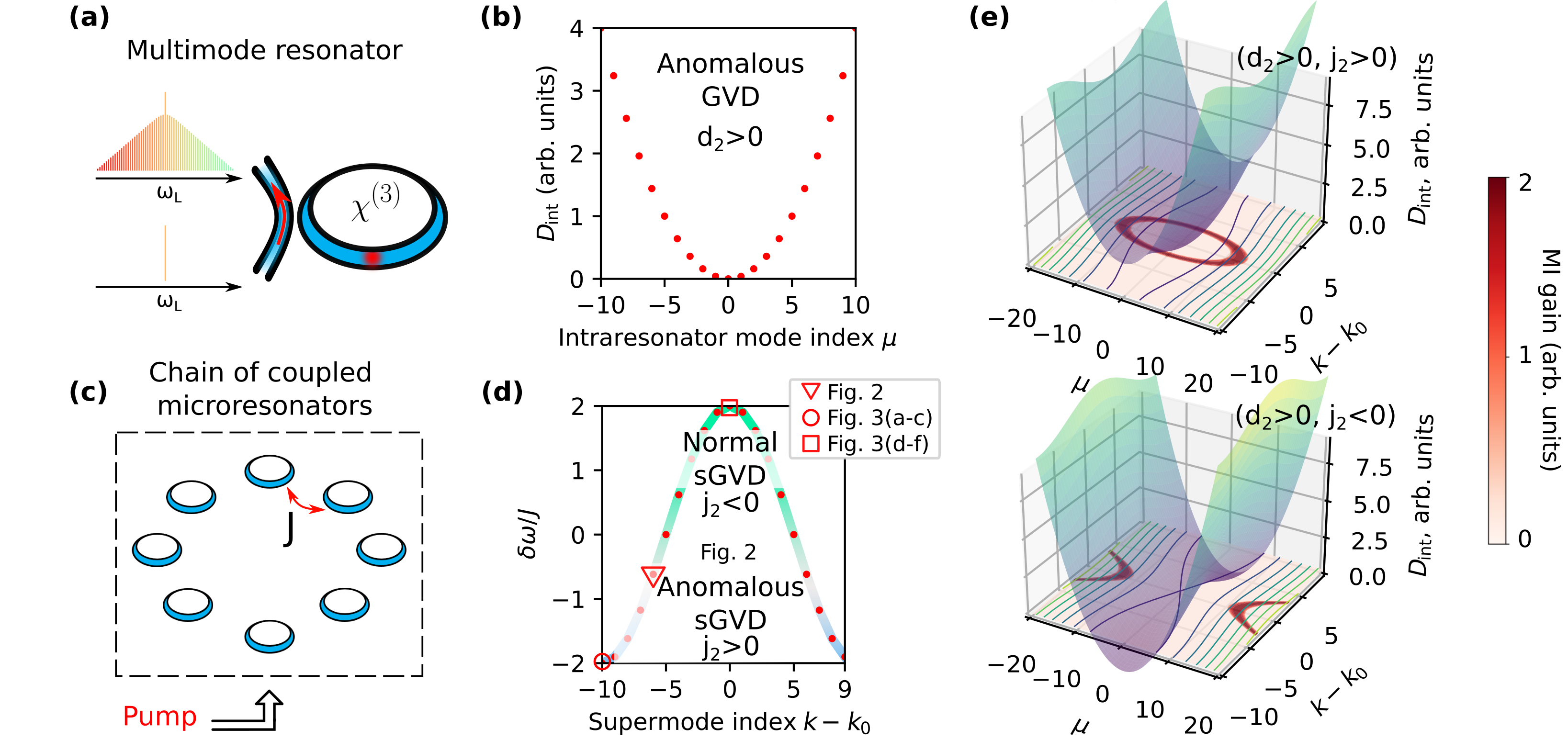}
\caption{\textbf{Hybridized dispersion in 1D lattice of equally coupled optical resonators}. 
    (a) Schematics of a continuous laser driven single optical $\chi^{(3)}$ resonator leading to dissipative Kerr soliton formation (i.e. an optical frequency comb); 
    (b) corresponding integrated dispersion profile which includes second-order dispersion $d_2$. (c) Schematics of 1D lattice in the ring configuration. The resonators are coupled with a rate $J$; (d) corresponding cosine band structure in the single-mode case with respect to $k_0=N/2$. Normal ($j_2<0$) and anomalous ($j_2>0$) regions of the band structure are depicted by green and blue, respectively. Triangle, rectangle, and circle indicate the pumped supermodes for generation of a traveling
    soliton in Fig.~\ref{fig:pic2} and investigation of the regularized wave collapse in Fig.~\ref{fig:pic3}. (e) Hybridized integrated dispersion of a multi-mode chain and modulation instability gain lobes (depicted in red) in elliptic (top panel, anomalous sGVD at $k_0=0$)  and hyperbolic (bottom panel, normal sGVD at $k_0=N/2$) regions, respectively.}% In plot (f), k is shifted by 10 for clarity.}
    \label{fig:pic1}
\end{figure*}

\section{Introduction}
 Over the past decade, it has been shown that continuous wave-driven Kerr nonlinear resonators host a variety of coherent dissipative structures~\cite{lugiato2018LugiatoLefeverEquation}~\cite{Kippenberg2018Dissipative}. In the anomalous dispersion regime, they give rise to dissipative Kerr solitons~\cite{Herr2014Temporal}, while in the normal dispersion regime, platicons~\cite{lobanov2015frequency,xue2015mode}, or interlocked switching
 waves, have been generated. These coherent dissipative
 structures give rise to a wide range of nonlinear dynamical phenomena, ranging from breathers~\cite{weiner2016observation, yu2017breather, lucas2017BreathingDissipativeSolitons} and soliton switching~\cite{guo2017UniversalDynamicsDeterministic} to chaotic behavior\cite{anderson2016observations}. Mathematically, in leading order, the dynamics can be described by the 1D driven-dissipative nonlinear Schr\"odinger equation (NLSE)~\cite{copie2020physics} known as the Lugiato-Lefever equation
 (LLE)~\cite{lugiato2018lugiato,godey2014stability}, and extension thereof, e.g., to include multi-mode dynamics or the Raman nonlinearity~\cite{yi2017single}. 
 In this framework, a variety of nonlinear phenomena have been observed~\cite{qi2019dissipative,xue2015mode,lobanov2015frequency,Lucas2017Breathing,karpov2019dynamics,cherenkov2017dissipative,Skryabin2021Threshold}.

On the application side, in particular, the dissipative Kerr soliton formation process has been utilized and has enabled photonic integrated microresonator-based optical frequency comb generation (Fig.~\ref{fig:pic1}(a)) with applications ranging from coherent communications~\cite{Marin-Palomo2017Microresonator} and neuromorphic computing~\cite{Feldmann2021Parallel} to atomic clocks~\cite{papp2014microresonator}.

Yet to date, almost all experimental and theoretical works on `dissipative structures' in optically driven Kerr nonlinear resonators (be it fiber~\cite{Leo2013Instabilities,leo2010temporal} or microresonator based) have focused on the single resonator case, and only recently extended to the dimer case~\cite{tikan2021emergent, komagata2021dissipative, helgason2021dissipative}.
The recent advances in ultra-low loss nonlinear integrated platforms, particularly silicon nitride~\cite{ji2021methods,liu2021high}, have dramatically reduced the threshold for optical parametric oscillations and concomitant dissipative structure generation --- at and below the $\mathcal{O} ( \mathrm{mW} ) $ level. This indicates that large-scale arrays of coupled Kerr nonlinear resonators that combine spatial and synthetic frequency dimensions~\cite{HuRealization2020} are within
experimental reach --- yet their nonlinear dynamics under continuous-wave driving remain largely unexplored. Such systems are expected to exhibit rich and unexpected nonlinear dynamics.
Even the simple case of a photonic dimer has demonstrated a variety of emergent nonlinear dynamics~\cite{tikan2021emergent,komagata2021dissipative} and novel phenomena such as soliton hopping and recurrent dispersive waves, as well as other surprising features in the stability chart, such as allowing DKS formation with higher efficiency. 1D and 2D lattices are particularly attractive as they allow significantly more complex dispersion landscapes --- opening new ways to engineer dispersion in
ways that are inaccessible using traditional dispersion engineering approaches. Therefore chains of resonators are expected to provide a pathway to octave-spanning dissipative Kerr solitons~\cite{li2017StablyAccessingOctavespanning}, which is an enduring outstanding challenge in the field. Such spectra are required for self-referencing of micro-combs~\cite{cundiff2003ColloquiumFemtosecondOpticala}. 
Moreover, they allow exploring 1D and 2D topological band structures by using a staggered coupling, such as the Su-Schrieffer-Heeger (SSH) model, or honeycomb strained graphene photonic lattices~\cite{ozawa2019topological,youssefi2021superconducting}. However, to date, the dissipative Kerr soliton formation in systems that exhibit bands, i.e., arbitrary lattices of 1D or 2D resonators, are not studied to the best of our knowledge.  One exception is the recently investigated~\cite{mittal2021topological} Kerr nonlinear version of the photonic 2D Haldane model made of coupled multi-mode optical microresonators with anomalous dispersion (Fig.~\ref{fig:pic1}(b)) that are coupled via link resonators.
Numerical simulations of this system demonstrated the formation of 'nested' dissipative Kerr solitons in the edge state, which is composed of individual DKS on each resonator site (which are continuous along the angular coordinate), whose amplitude along the edge state constitutes a propagating waveform (i.e., a discrete soliton).  
For this reason, the system has two coherent timescales: the round trip cavity time and the overall propagation time along the edge. However, this work left several key aspects not answered.
Specifically, under what condition can such two-dimensional (i.e., in spatial (sparse) and synthetic frequency (dense) dimensions) dissipative structures be coherent? How does the extension of dimensionality change the four-wave mixing (FWM) processes in general and soliton (or more generally, coherent dissipative structures) generation in particular? Moreover, an open question is to what extent topologically protected (i.e., robust) edge states~\cite{Asboth2016Short} (be it 2D Haldane, 1D SSH or any other model) can give rise to coherent solitons, and to what extent soliton formation is robust~\cite{tikan2021emergent,mittal2021topological}. Crucially, the description of the interaction of the edge state with bulk mode has not been investigated to date. This could explain the unanswered observation of partially coherent dynamics of the recently numerically predicted edge solitons in the Haldane lattice of nonlinear coupled resonators~\cite{mittal2021topological}.

Here we theoretically analyze nonlinear dynamics in an arbitrary lattice of coupled resonators and study in detail the dynamics of soliton formation in a 1D chain (with periodic boundary conditions, cf. Fig.~\ref{fig:pic1}(c,d)). We show that the system exhibits a rich 2D spatio-temporal nonlinear dynamics. Specifically, we demonstrate the formation of coherent spatio-temporal solitons in equally coupled, topologically trivial, 1D chains and
demonstrate that the mean-field model describing the dynamics can be represented in the form of elliptic or hyperbolic 2D LLE (cf. Fig.~\ref{fig:pic1}(e))~\cite{Ablowitz2021Transverse}, thereby making a link to the prior findings in the field of spatial solitons~\cite{SCROGGIE19941323,Firth2002Dynamical} and  generalizing the conventional single-mode coupled-resonator theory~\cite{Morichetti2012First,marti2020slow}. We describe the principles of 2D FWM processes and the global soliton
formation, showing that the system is equivalent to the solutions found to exist in the edge state of the 2D topological Haldane model~\cite{mittal2021topological}. We also predict novel emergent nonlinear effects such as edge-to-bulk scattering and regularized wave collapse. 
Equally,  we identify the breaking down mechanism of the topological edge state formed soliton using the simplest example of the 1D SSH topological model due to the more complex four wave mixing pathways. Our work bridges the knowledge gap between the simplest and relatively well-understood case of the photonic dimer~\cite{tikan2020symmetry,komagata2021dissipative} and the topological arrangement studied in Ref.~\cite{mittal2021topological}, revealing the dramatic change of the dynamics caused by the increased dimensionality of the system.

\section{Coupled Lugiato-Lefever equations in lattices of resonators}

We start with a general description of a system of weakly coupled identical optical resonators that is shown to be governed by a set of linearly coupled LLEs, which can be presented in matrix form as 
\begin{equation}\label{eq:LLE_lattice}
    \frac{\partial}{\partial t}\mathbb{A} = \hat{\mathbb{D}}\mathbb{A} +i\hat{\mathbb{M}}\mathbb{A} + ig_0|\mathbb{A}|^2\mathbb{A} + \mathbb{F},
\end{equation}
where vector $\mathbb{A} = [A_0, ..., A_{N-1}]^T$ contains optical field envelopes of each resonator in the lattice, matrix 
\begin{align*}
    \hat{\mathbb{D}} &= \mathrm{diag}\big[ -\big(\frac{\kappa_{0}+\kappa_{\mathrm{ex},0}}{2}+i\delta\omega_0\big) + i\frac{D_2}{2}\frac{\partial^2}{\partial\varphi^2}, ...,\\
    &...,-\big(\frac{\kappa_{0}+\kappa_{\mathrm{ex},N-1}}{2}+i\delta\omega_0\big) + i\frac{D_2}{2}\frac{\partial^2}{\partial\varphi^2}  \big]
\end{align*}
contains detuning ($\delta \omega_0$), losses ($\kappa_0$), dispersion of each resonator ($D_2$), and coupling to the bus waveguides ($\kappa_{\mathrm{ex},\ell}$). The coupling between different rings is introduced in matrix $\hat{\mathbb{M}}$, the nonlinear term $|\mathbb{A}|^2\mathbb{A} = [|A_0|^2 A_0,..., |A_{N-1}|^2 A_{N-1}]^T $ describes the conventional Kerr nonlinearity with single photon Kerr frequency shift $g_0$, and $\mathbb{F} = [\sqrt{\kappa_{\mathrm{ex},0}} s_{\mathrm{in},0},..., \sqrt{\kappa_{\mathrm{ex},N-1}}s_{\mathrm{ex},N-1}]^T$ represents
the pump. Usually, the coupling matrix $\hat{\mathbb{M}}$ is diagonalizable and possesses a set of eigenvectors $\{\mathbb{V}_i\}$ and associated eigenvalues $\lambda_i$, so any state $\mathbb{A}$ can be represented in this basis
\begin{equation}
    \mathbb{A} = \sum\limits_jc_j \mathbb{V}_j,
\end{equation}
where coefficients $c_j = \braket{\mathbb{A}|\mathbb{V}_j}$ correspond to the amplitude of the collective mode $\mathbb{V}_j$  and $\braket{\cdot|\cdot}$ indicates scalar product. Therefore, Eq.~(\ref{eq:LLE_lattice}) can be rewritten for the amplitudes $c_j$ in the basis of eigenvectors $\{\mathbb{V}_j\}$, where the the linear part of the equation will take a form of a matrix with eigenvalues $\lambda_i$ on the diagonals corresponding to the resonance frequencies of the collective excitations. However, the nonlinear term will be no longer diagonal in this basis. In the direct space, the nonlinear term takes form
\begin{equation*}
    |\mathbb{A}|^2 \mathbb{A} = \sum\limits_{j_1,j_2,j_3} c_{j_1} c_{j_2}c_{j_3}^*\mathbb{V}_{j_1} \mathbb{V}_{j_2} \mathbb{V}_{j_3}^*.
\end{equation*}
Projecting this expression onto the state $\mathbb{V}_j$, one obtains the coupled-mode equations for the amplitudes $c_j$
\begin{align}
    \frac{\partial c_j}{\partial t} &= -(\frac{\kappa_0 + \kappa_\mathrm{ex}}{2} + i(\delta\omega_0 - \lambda_j) )c_j + i\frac{D_2}{2}\frac{\partial^2 c_j}{\partial \varphi^2} + \nonumber \\
    &+ig_0\sum\limits_{j_1,j_2,j_3}c_{j_1}c_{j_2}c_{j_3}^*\braket{\mathbb{V}_{j_1} \mathbb{V}_{j_2} \mathbb{V}_{j_3}^*|\mathbb{V}_{j}} + \tilde{f}_j, 
\end{align}

where $\tilde{f}_j = \braket{\mathbb{F}|\mathbb{V}_j}$ is projection of the pump on the eigenstate $\mathbb{V}_j$, the nonlinear term represents the conventional four-wave mixing process with the conservation law dictated by the product $\braket{\mathbb{V}_{j_1} \mathbb{V}_{j_2} \mathbb{V}_{j_3}^*|\mathbb{V}_{j}}$. The eigenvalues $\lambda_j$, showing the dependence of supermode frequency on supermode number, naturally start to play a role of dispersion, similar to the conventional LLE in a single resonator. In general, the eigenvalues $\lambda_j$ are not equidistantly separated, and the supermode dispersion can be introduced like the integrated dispersion of a single resonator $D_\mathrm{int}(k) = \lambda_k - k J_1(k-k_0)$, where $J_1$ is the local free spectral range of the spatial supermodes in the vicinity of $k_0$.

\begin{figure*}
    
    \centering
    \includegraphics[width=\textwidth]{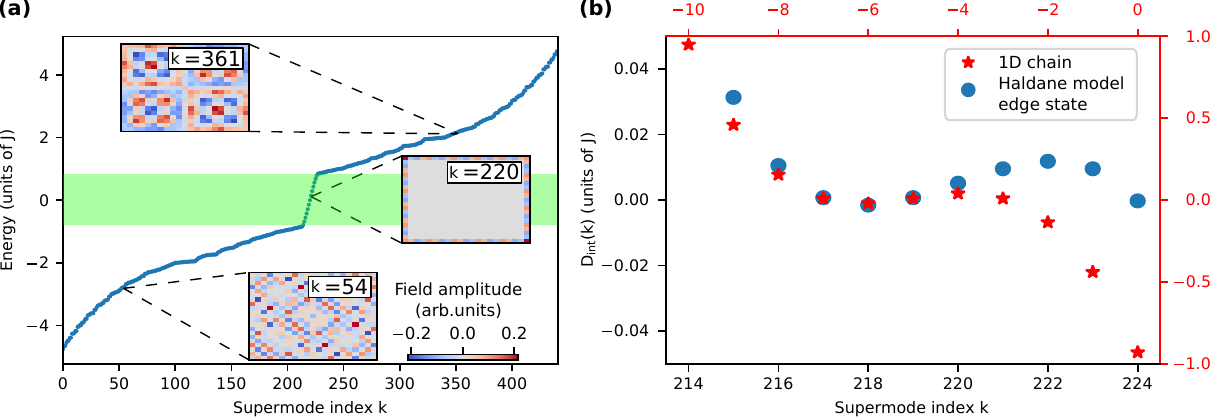}
    \caption{\textbf{Band structure of a photonic Haldane model.} Band structure of a Haldane lattice consisting of $21\times 21$ optical resonators is shown in panel (a). The edge states are highlighted by the green rectangle. Insets represent the field amplitude in each resonator for the eigenstates $k=54,\,220,\,361$. Panel (b) shows integrated supermode dispersion $D_\mathrm{int}(k)$ for the edge state of the Haldane model (blue dots) and the chain of 20 equally coupled microresonators (red stars). }
    \label{fig:SI_Haldane}
\end{figure*}

Furthermore, this reasoning can be applied to coupled system with non-trivial topologies, including the Haldane model considered in Ref.~\cite{mittal2021topological} with the lattice of $21 \times 21$ resonators. Diagonalization of the coupling matrix $\mathbb{M}$ yields the band structure (shown in Fig.~\ref{fig:SI_Haldane}(a)) with three remarkable regions: lower bulk, upper bulk, and edge states. The integrated dispersion for the edge modes, shown by blue dots in Fig.~\ref{fig:SI_Haldane}(b), reveals a typical dispersion curve with pronounced second and third-order dispersion coefficients. Pumping a given supermode $k$ above a given threshold, four-wave mixing processes can occur and lead to generation of frequency combs, which dynamics and bandwidth will be determined by the local dispersion profile; therefore, the excitation of the supermode from the center of the edge band will be mainly determined by the neighbouring edge supermodes.  Remarkably, a chain of $20$ equally coupled resonators (depicted by the red stars in Fig.~\ref{fig:SI_Haldane}(b)) has a similar profile of supermode dispersion. Neglecting the presence of the bulk modes in the Haldane lattice, the nonlinear dynamics of the edge states can be modelled as a simple chain of resonators. The simplified model provides an opportunity to analytically investigate general aspects of the dynamics and find analogies with already known effects in nonlinear physics.

\section{Chains of coupled microresonators}
\subsection{Two-dimensional hybridized dispersion}
We continue our analysis by considering a system of equally coupled chain of resonators. Dynamics of the optical field envelope $A_\ell$ in $\ell$-th resonator is described by the following system of coupled LLEs
\begin{align}\label{eq:General_couped_LLEs}
    \frac{\partial A_\ell}{\partial t} &= -(\frac{\kappa_{\mathrm{ex},\ell} + \kappa_0}{2} + i \delta\omega_0)A_\ell + i J(A_{\ell-1}+ A_{\ell+1})\nonumber\\
    +& i\frac{D_2}{2}\frac{\partial^2 A_\ell}{\partial \varphi^2} + i g_0 |A_\ell|^2 A_\ell +\sqrt{\kappa_{\mathrm{ex},\ell}}s_{\mathrm{in},\ell} e^{i \phi_\ell}.
\end{align}
Here $\kappa_{\mathrm{ex},\ell}$ is the coupling of the $\ell$-th resonator to the corresponding pump $s_{\mathrm{in},\ell}$ with its general phase $\phi_\ell$, $\omega_0$ is laser-cavity detuning, $\kappa_0$ is intrinsic linewidth of the resonators, $J$ is coupling strength between the neighbouring resonators, $D_2$ is chromatic group velocity dispersion (GVD). 
For simplicity, in the case of constant couplings to the bus waveguides $\kappa_{\mathrm{ex},\ell}$ and constant inter-resonator couplings $J$, we introduce normalized variables $d_2 = D_2/\kappa$, $\kappa = \kappa_0 +\kappa_\mathrm{ex}$, $\zeta_0=2\delta\omega/\kappa$, $j = 2J/\kappa$, $f_\ell = \sqrt{8\kappa_\mathrm{ex}g_0/\kappa^3}s_{\mathrm{in},\ell} e^{i\phi_\ell}$,  $\Psi_\ell = \sqrt{2g_0/\kappa}A_\ell$. In the normalized units, Eq.~(\ref{eq:General_couped_LLEs}) reads
\begin{equation}\label{SIeq:normalized_coupled_LLEs}
     \frac{\partial \Psi_\ell}{\partial \tau} = -(1+i\zeta_0)\Psi_\ell + id_2\frac{\partial^2 \Psi_\ell}{\partial \varphi^2}+ ij\big(\Psi_{\ell-1}+\Psi_{\ell+1} \big) + i|\Psi_\ell|^2 \Psi_\ell + f_\ell.
\end{equation}
The linear part can be diagonalized by the Fourier transform
\begin{equation}\label{SIeq:Fourier_transform}
    \psi_{\mu k} = \frac{1}{2\pi\sqrt{N}}\int\sum_{\ell = 1}^N \Psi_\ell e^{2\pi i(\ell k/N +\mu\varphi)}d\varphi,
\end{equation}
where $k$ is the supermode index and $\mu$ is the comb line index. With the Kerr term, Eq.~(\ref{SIeq:normalized_coupled_LLEs}) transforms to
\begin{align}\label{SIeq:LLE_in_fourier}
    \frac{\partial \psi_{\mu k}}{\partial \tau} &= -(1+i\zeta_0)\psi_{\mu k} - i \big[d_2 \mu^2 - 2j\cos{\frac{2\pi k}{N}}]\psi_{\mu k}+ \delta_{\mu0}\tilde{f}_{k}+ \nonumber \\
     &+\frac{i}{N}\sum_{\substack{k_1,k_2,k_3\\\mu_1 \mu_2 \mu_3}}\psi_{\mu_1 k_1}\psi_{\mu_2 k_2}\psi_{\mu_3 k_3}^*\delta_{\mu_1+\mu_2-\mu_3-\mu}\delta_{k_1+k_2-k_3-k}.
\end{align}
The term in the square brackets is normalized integrated dispersion defined from Eq.~(\ref{eq:res_pos}) as $d_\mathrm{int}(\mu,k)=2(\omega_{\mu k}-\omega_0+D_1\mu)/\kappa$, and it  incorporates the dispersion laws for resonator ($d_2 \mu^2$) and supermodes ($2j\cos2\pi k/N$), representing the hybridized 2D dispersion surface 
\begin{equation}\label{eq:res_pos}
d_\mathrm{int}(\mu,k)=d_2 \mu^2 -2j\cos(2\pi k/N). 
\end{equation}
In the case of anomalous GVD ($d_2>0$, Fig.~\ref{fig:pic1}(b)) of the individual resonator,  this surface with parabolic and cosine cross-sections is shown in Fig.~\ref{fig:pic1}(e). Local dispersion topography changes along the $k$ axis, revealing different regions with parabolic and saddle shapes. This hybridization and the corresponding 2D dispersion surface apply to any lattice of resonators, including topologically non-trivial.
The pump term $\tilde{f}_k$ stands for the projection of the pump on the $k$-th supermode
\begin{equation}\label{SIeq:pump_projection}
    \tilde{f}_{k} = \frac{1}{\sqrt{N}}\sum_{\ell = 1}^N f_\ell e^{2\pi i\ell k/N }.
\end{equation}

\begin{figure*}
    \centering
    \includegraphics[width=\textwidth]{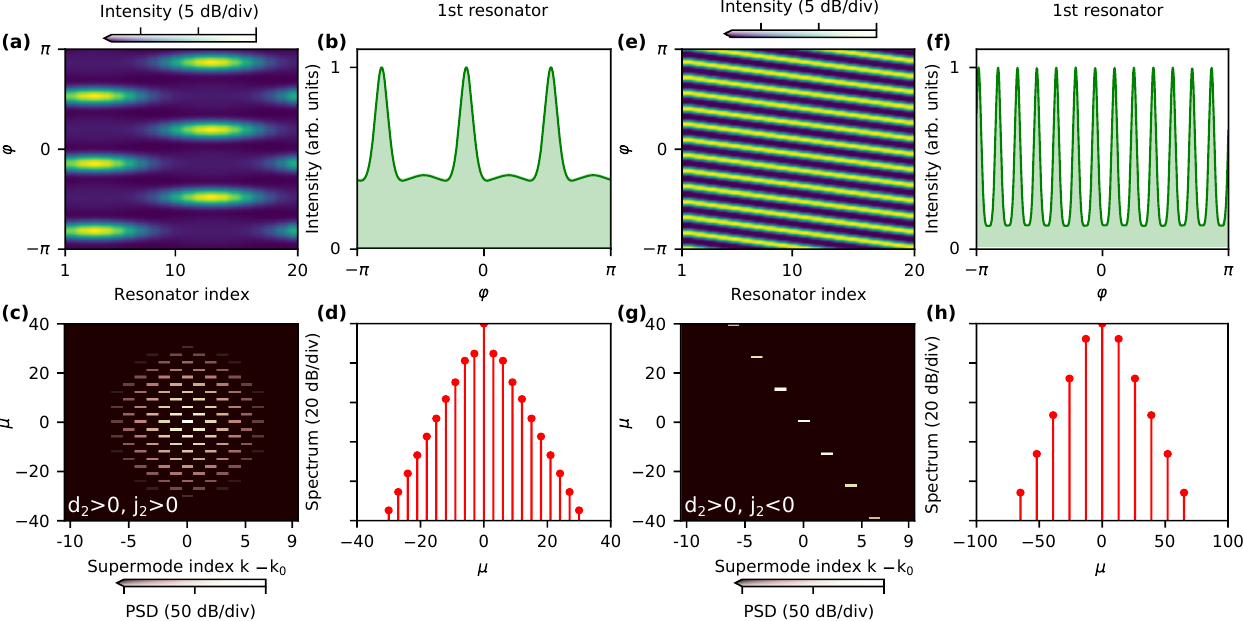}
    \caption{\textbf{Coherent dissipative structures in a driven nonlinear photonic ring lattice}. Panels (a-d) correspond to the elliptic region ($k_0 = 0$, $d_2>0,\, j_2>0$), and panels (e-h) correspond to the hyperbolic ($k_0 = N/2$, $d_2>0,\, j_2<0$). Spatio-temporal profiles of the mode-locked structures are shown in panels (a,e) with the corresponding field profile on a single resonator level in panels (b,f). The 2D spectral profiles of the states (a) and (e) obtained via Eq.~(\ref{SIeq:Fourier_transform}) are presented in (c) and (g), respectively. The spectral profile in elliptic regime (c) forms a disk, whereas the spectrum of the pattern in hyperbolic regime (g) tends to align one of the asymptotes of the hyperbola depicting modulation instability gain
    %in Eq.~(\ref{SIeq:MI_diag}). 
    The Fourier spectra of the states (b) and (f) are presented in (d) and (h).}
    \label{fig:SIpic2}
\end{figure*}

\subsection{Spatial eigenstates and pump projection on the chain}
With the assumptions above, we can consider the proposed structure as a perfect 1D photonic crystal that naturally possesses a set of collective spatial excitations or \emph{supermodes} whose eigenvalues form a cosine band structure schematically shown in Fig.~\ref{fig:pic1}(d).  The band structure describes the energy range of the excitations propagating in the crystal and imposes their dispersion law, which plays a crucial role in the context of nonlinear physics. The regions of anomalous and normal supermode GVD (sGVD) are shown in Fig.~\ref{fig:pic1}(d) by blue and green colors, respectively. For a given supermode index $k_0$, the linear term in the Taylor series gives the supermode FSR equal to $J_1/2\pi = 2J/N \sin(2\pi k_0/N)$ and the corresponding quadratic term yields sGVD $J_2=2J(2\pi/N)^2\cos(2\pi k_0/N)$.

The excitation of the individual supermode requires an accurate pump projection on its spatial profile. Though the excitation of the system via a single resonator is possible, the pump power, in this case, will be redistributed among all the supermodes within the excitation bandwidth. The number of the excited modes will depend on the local density of states within the width of the band in Fig.~\ref{fig:pic1}(d). Even if the resonance linewidth is small enough, so the individual resonances in Fig.~\ref{fig:pic1}(d) are distinguishable, the single-resonator pump scheme always leads to the excitation of supermodes in pairs due to their two-fold degeneracy, except for the modes from the very top and bottom of the band. For simplicity, in the following we focus on the ideal case of a single supermode excitation.
The pump efficiency and the number of excited supermodes in the chain of resonators depend on the spatial arrangement of the pump scheme and the density of states of supermodes.
According to Eq.~(\ref{SIeq:pump_projection}),
if the resonator $\ell=0$ is pumped,  all the supermodes have a pump term with the projection amplitude $1/\sqrt{N}$. With the increasing number of resonators, pumping scheme with a single resonator excitation becomes less efficient, and more sophisticated schemes are required. 
To excite only one supermode with index $k_0$, one needs to adjust the relative phases of the pump lasers accurately; thus, the maximal projection on the supermode $k_0$ will be achieved for pump configuration
\begin{equation} \label{SIeq:pump_k}
    \mathbf{f} = f^{(0)}\big[1,e^{-2\pi i k_0/N},e^{-4\pi i k_0/N},...,e^{-2(N-1)\pi i k_0/N}  \big],
\end{equation}
where $f^{(0)} = \sqrt{8 g_0 \kappa_\mathrm{ex}P/\kappa^3\hbar\omega N}$ is normalized pump for a single resonator.

\subsection{Modulation instability gain lobes.}
Further, we investigate the stability of plane wave solutions $\psi_{00}$.
Considering the pump at $\mu_0=0$ and $k-k_0=0$ ($k-k_0=N/2$), we investigate FWM processes between the pump mode and the modes with indexes $\mu,k$. Linearizing the system with respect to these modes, we identify the modes with positive parametric gain. Our analysis, similar to Ref.~\cite{Ablowitz2021Transverse}, shows that the modulationally unstable solutions form an ellipse (hyperbola) in the $\mu-k$ space.
\begin{equation}\label{eq:MI_diag}
    d_2 \mu^2 \pm j_2 k^2 = 4|\psi_{00}|^4 + \sqrt{|\psi_{00}|^4-1}-(\zeta_0\mp 2j),
\end{equation}
here $+$ $(-)$ stands for the excitation of $k-k_0=0$ ($k-k_0=N/2$). An example of the modulation instability gain lobes [Eq.~(\ref{eq:MI_diag})] is presented in Fig.~\ref{fig:pic1}(e) for both regions in case of $d_2 = 0.04$ and $j_2 = 2 |J_2|/\kappa=1$. Top panel in Fig.~\ref{fig:pic1}(e) reveals that the supermode corresponding to the excitation of all the resonators in-phase (anomalous sGVD) is unstable against small perturbations with $\mu$ and $k$ indexes that form an ellipse.
The width and height of the ellipse are defined by pump power, $d_2$, and $j_2$ coefficients that correspond to GVD and sGVD. In contrast, the state corresponding to the excitation of the neighbouring resonators in the opposite phase (normal sGVD) is unstable against the perturbations with  $\mu$ and $k$ forming a hyperbola, showing that all the supermodes can experience positive parametric gain.

\begin{figure*}
    \centering
    \includegraphics[width=\textwidth]{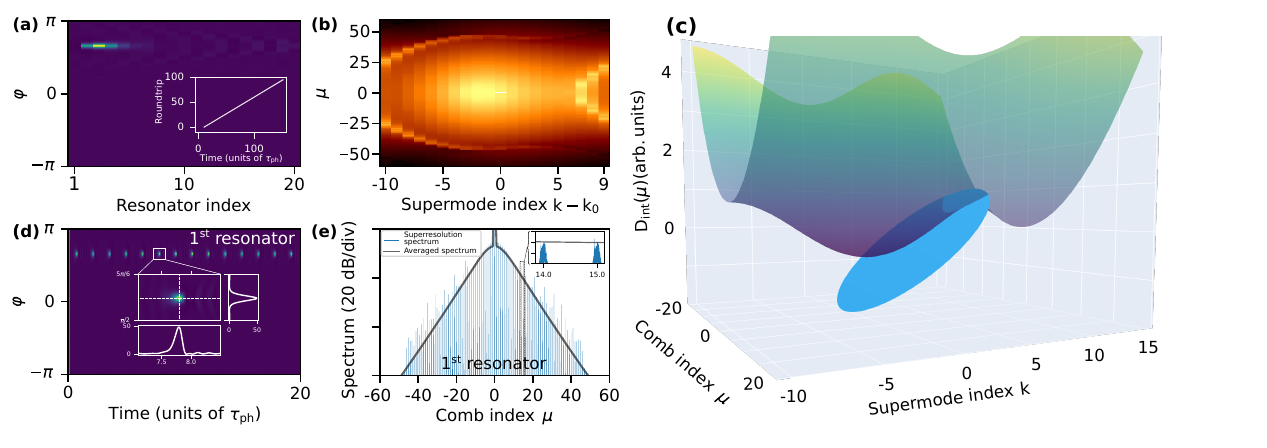}
    \caption{\textbf{Localized 2D dissipative soliton in a chain of 20 resonators.} Instantaneous field profile in the chain of resonators and the corresponding 2D spectral profile are shown in panels (a) and (b). Inset in (a) shows the roundtrip number of the soliton in time. Schematics of the soliton as a dispersionless structure beneath the hybridized dispersion surface is shown in panel (c).  Panel (d) represents the field dynamics on a single resonator level. The corresponding superresolution and averaged spectrum are presented in panel (e).}\label{fig:pic2}
\end{figure*}

\subsection{Coherent dissipative structures.}
We continue with the simulation of the coupled LLEs in Eq.~(\ref{eq:General_couped_LLEs}) for 20 resonator chain and constant normalized coupling $j = 10.13$ ($j_2=1$). To simulate the temporal dynamics, we employ the step-adaptative Dormand-Prince Runge-Kutta method of Order 8(5,3)~\cite{press2007numerical} and approximate the dispersion operator by the second-order finite difference scheme. We deliberately choose the pumping scheme allowing for exciting only a given mode. To trigger the FWM processes, we numerically scan the resonance with a fixed pump power and track field dynamics in all the resonators.
\paragraph{Turing patterns}
We simulate nonlinear dynamics and Turing patterns in hyperbolic $k_0=N/2$ and elliptic $k_0=0$ regimes. 
To observe coherent structures, we scan the resonance by changing the normalized laser detuning $\zeta_0$ and bring the system into an unstable state. Having stimulated the pattern formation, we further tune towards the monostable region ($\zeta_0^{k_0}=\zeta_0\mp 2j<\sqrt{3}$, $+$ ($-$) stands for $k_0 = 0$ ($N/2$)) and obtain stable coherent structures in both regimes (Fig.~\ref{fig:SIpic2}). One can see that in the elliptic regime at $|f_\ell| = 1.05$ and $\zeta_0 = 20.5 $, we observe the formation of a hexagonal pattern [Fig.~\ref{fig:SIpic2}(a)]~\cite{SCROGGIE19941323,PhysRevLett.66.2597,Firth1992Hexagonal}. On a single resonator level, this corresponds to locked pulses [Fig.~\ref{fig:SIpic2}(b)] with a typical comb spectrum shown in Fig.~\ref{fig:SIpic2}(d). The corresponding 2D $k$-$\mu$ spectral profile in Fig.~\ref{fig:SIpic2}(c) shows that the sidebands form a disk, occupying the supermodes from both anomalous ($|k-k_0|<5$) and normal dispersion regimes. In the hyperbolic regime at $|f_\ell| = 2.35$ and $\zeta_0 = -20.3$, we observe a train of pulses in each resonator locked to each other [Fig.~\ref{fig:SIpic2}(e,f)]. The corresponding 2D spectral profile [Fig.~\ref{fig:pic2}(g)] forms a line in $k$-$\mu$ space, that qualitatively follows one of the asymptotes of the hyperbola that depicts modulation instability gain lobes in Fig.~\ref{fig:pic1}(e). Comparing the comb spectra at the 1st resonator Fig.~\ref{fig:pic2}(d)] with the elliptic case [Fig.~\ref{fig:pic2}(h)], one can notice that the state at the hyperbolic regime has a wider comb spectrum.

\paragraph{Spatio-temporal two-dimensional dissipative soliton}
We also generate a localized 2D dissipative solitons~\cite{Firth2002Dynamical} traveling along the circumference of the chain, which we describe in the following. To generate this spatio-temporal Kerr soliton (2D-DKS), we pump the 4th supermode in the elliptic regime ($k-k_0=-6$ marked by the red triangle in Fig.~\ref{fig:pic1}(d)) with $|f_\ell|=2.35$ and $\zeta_0 = 10.92$, so the local dispersion has anomalous sGVD $j_2 = 4j(\pi/10)^2\cos2\pi/5$ in addition to the non-zero supermode FSR
$j_1 = 2j\pi/5\sin 2\pi/5$. The obtained solution of the 2D-DKS corresponds to continuously re-circulating spatial discrete soliton that forms an ellipse with a fish-like tail in the spectral domain (cf. Fig.~\ref{fig:pic2}(a,b)). Similar to Cherenkov radiation for conventional DKS, the disk-shaped soliton crosses the hybridized dispersion in the vicinity of the edge of the Brillouin zone (cf. Fig.~\ref{fig:pic2}(c)),  resulting in the intensive generation of the dispersive waves, forming the fish-like spectrum, but presuming the soliton coherence. On the single resonator level, the optical field envelope demonstrates breathing dynamics (Fig.~\ref{fig:pic2}(d)) because the pulses periodically arrive in the resonator. Resolving the field envelope dynamics in time, one detects the periodic appearance of optical pulses and adjacent dispersive waves. Sampling this signal in time and computing the overall Fourier transform gives the so-called superresolution spectrum shown in Fig.~\ref{fig:pic2}(e). The periodic nature of the signal reveals a typical comb spectrum, with the presence of a fine spectral structure around each comb line, shown in the inset of Fig.~\ref{fig:pic2}(e). These subcombs appear due to the breathing dynamics and emergence of the corresponding dispersive waves, and the number of the spatial modes does not define the number of these subcombs. In fact, these subcombs correspond to just low frequency breathing, which is also present in the single resonator case in the breathing regime~\cite{Lucas2017Breathing}.  The time-averaging of the signal yields a smooth spectral profile (solid line in Fig.~\ref{fig:pic2}(e)), indicating the periodic nature of the signal. Notably, a similar (in terms of hybridized dispersion) 2D-DKS was observed in the edge state of the Haldane model~\cite{mittal2021topological}. However, due to the presence of the other bands, FWM-induced edge-to-bulk scattering strongly influenced soliton stability, resulting in temporal decoherence. This effect can be understood via considering a DKS generated at the edge state of the SSH chain~\cite{Asboth2016Short} described in the following.

\begin{figure*}
    
    \centering
    \includegraphics[width=\textwidth]{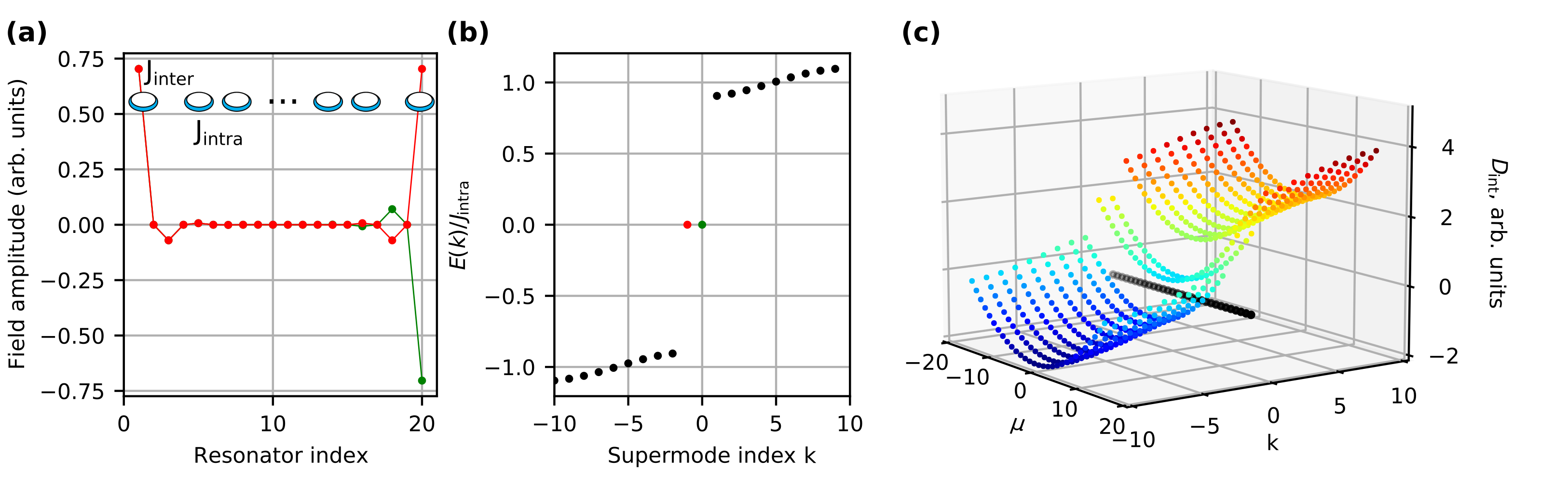}
    \caption{\textbf{Dissipative Kerr solitons at the edge state of the Su-Schrieffer--Heeger model (SSH).} Panel (a) represents the spatial profile of the two edge states of the Su-Schrieffer--Heeger model (SSH) model consisting of $20$ optical microresonators with the schematics of the chain in the inset. Band structure of the SSH chain of 20 resonators is shown in panel (b). The hybridized dispersion profile and schematics of the generated soliton  at the edge state (black line below the edge state parabola) are shown in panel (c).  }
    \label{fig:SI_SSH}
\end{figure*}

\subsection{Nonlinearly induced edge-to-bulk scattering in the Su-Schrieffer--Heeger model}
The edge states of the SSH model are localized on the corners of the chain as shown in Fig.~\ref{fig:SI_SSH}(a). The chain supports edge states in the case where inter-cell coupling $J_\mathrm{inter}$ is smaller than intra-cell coupling $J_\mathrm{intra}$ (also shown in the inset in Fig.~\ref{fig:SI_SSH}(a)). In the limit $J_\mathrm{inter}\to 0$ (trivial edge state~\cite{Asboth2016Short}), the first resonator is completely decoupled from the chain, and its dynamics is described by conventional LLE. With the finite ratio $J_\mathrm{inter}/J_\mathrm{intra}<1$ the formed band structure (see Fig.~\ref{fig:SI_SSH}(b)) has upper and lower bulk regions with eigenmodes in the middle of the gap that correspond to the edge states~\cite{Tusnin2021Dissipative}. With the chromatic dispersion taken into account, the nonlinear interactions happen on the hybridized dispersion surface, presented in Fig.~\ref{fig:SI_SSH}(c). Generation of the edge soliton corresponds to the formation of the dispersionless line below the edge state parabola (schematically shown in Fig.~\ref{fig:SI_SSH}(c)). If the width of the bandgap is large enough (effectively corresponds to limit $J_\mathrm{inter}/J_\mathrm{intra}\to 0$, $J_\mathrm{intra}\gg \kappa$), the dynamics of the soliton will be similar to the single-resonator dynamics, because the field will be still localized in the first ring. However, if the soliton line can cross the lower bulk band, additional photon transfer to the bulk modes will occur (similar effect has already been observed in the system of just two coupled resonators considered in Ref.~\cite{tikan2021emergent}). The photons scattered to the bulk will experience now two-dimensional dynamics and drastically affect the soliton stability. While we leave the accurate description for future work, similar effect has been observed experimentally in dimers~\cite{tikan2021emergent,komagata2021dissipative} and trimers~\cite{tikan2020symmetry}; therefore, we can extend our qualitative analysis to higher-dimensional topological models. For example, in a 2D lattice, a 2D soliton generated at 1D edge state will scattered to the edge, where the dispersive waves will experience 3D nonlinear dynamics. If a corner state is realized in a 2D lattice, the corresponding corner state soliton will be similar to the conventional single-resonator DKS, however scattering to bulk will lead to 3D nonlinear dynamics of the bulk as well.

\begin{figure*}
    \centering
    \includegraphics[width=\textwidth]{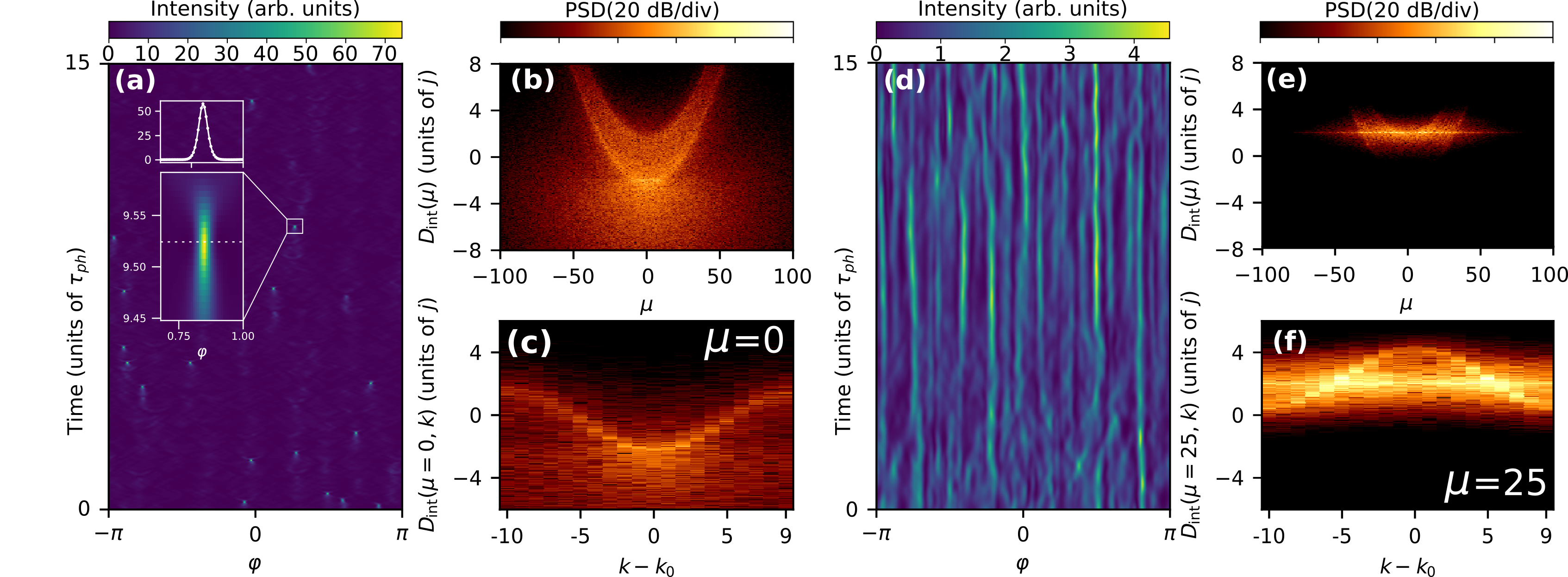}
    \caption{\textbf{Numerical reconstruction of the nonlinear dispersion relation in the elliptic and hyperbolic regions in the unstable regime.} Panels (a-c) correspond to the elliptic region ($k_0 = 0$, $d_2>0,\, j_2>0$), (b-f) to the hyperbolic ($k_0 = N/2$, $d_2>0,\, j_2<0$).  Spatiotemporal diagrams of unstable states in $0$th resonator are shown in (a) and (d); The corresponding nonlinear dispersion relation (NDR) in the elliptic region (b) demonstrates excitation of all the optical and spatial modes, whereas the NDR in the hyperbolic region (e) reveals that photon transfer between the spatial supermodes is suppressed in the vicinity of the pump mode $\mu=0$ ; The panes (c) and (f) represent the nonlinear supermode dispersion relation [Eq.~(\ref{eq:NDR})] of $0$th comb line for the state in (a) and $25$th comb line for the state in (d).}
    \label{fig:pic3}
\end{figure*}

\subsection{Wave collapse}
Since the LLE is the NLSE with an external forcing term and dissipation, it can possess similar features, and in particular, the effect called wave collapse~\cite{Zakharov1988Wave,Kartashov18Self}. Wave collapses play an important role in physics. In the conservative 2D NLSE, it reveals a singularity of the model, related to a possibility of full pulse compression in a finite time. Practically, this leads to an effective mechanism of local energy dissipation. 
It has been shown for 2D elliptic focusing NLSE that a pulse of a finite width can implode to an infinitely small area concentrating there a finite amount of energy~\cite{Kuznetsov2009SelfFocusing,Rasmussen_1986BlowUp} and therefore becoming ultra broad in the spectral domain. Even the presence of dissipation in 2D LLE does not restrict the wave collapses~\cite{tsutsumi1984nonexistence}. On the contrary, wave collapses do not occur in the 2D focusing NLSE with hyperbolic dispersion~\cite{Rasmussen_1986BlowUp}, signifying that it is the dispersion curvature  that is responsible for the effect. When wave collapses happen in real systems, the corresponding spectra become too large, so the simplest approximation with second-order dispersion operator becomes not valid anymore, and higher dispersion orders must be taken into account. Consequently, the pulse width does not completely compress, and the collapse regularizes~\cite{ilan2002self}. The same effect we observe in our model. Exciting incoherent dynamics by pumping the elliptic region at $|f_\ell|=2.35$ and $\zeta_0 = 22.1$, we observe rapid formation and dissipation of narrow pulses in each cavity. A typical spatio-temporal diagram at a single resonator level is shown in Fig.~\ref{fig:pic3}(a). 
We observe the random appearance of the pulses in different parts of the cavity and further their rapid compression, during which the peak amplitude significantly exceeds the background level. However, investigating the pulse width dynamics, we find that it does not completely shrink. To find what limits the minimum pulse width, we computed the nonlinear dispersion relation (NDR)~\cite{Leisman2019Effective,tikan2021emergent,tikan2021nonlinear} [Fig.~\ref{fig:pic3}(b)] that is the 2D Fourier transform of the spatio-temporal diagram of the complex field envelope in Fig.~\ref{fig:pic3}(a).
We observe the high photon occupancy of the region beneath the parabolas, which indicates the presence of 2D dissipative nonlinear structures. Furthermore, all the hybridized parabolas are populated by the photons, meaning that supermodes from both dispersion regions are excited.
We continue by reconstructing the supermode NDR for 0th comb line $(\mu_0=0)$ for \emph{all} resonators in the following way
\begin{equation}\label{eq:NDR}
    NDR(\Omega,\mu_0,k) = \frac{1}{\sqrt{N_t N}}\sum_{\ell,n}\psi_{\mu_0\ell}(t)e^{i (2 \pi k \ell/N - \Omega t_n)},
\end{equation} where $\Omega$ is slow frequency, $t_n = \Delta t n$ with $\Delta t = T/N_t$ time-step, $T$ is simulation time with $N_t$ number of discretization points. The result is shown in Fig.~\ref{fig:pic3}(c). The whole cosine band structure is populated, including the region of the normal dispersion that prevents the full wave collapse. 

We continue the analysis by exciting the hyperbolic region under the same conditions (same pump power and relative detuning $\zeta_0 = -17.0$). As mentioned earlier, the local dispersion topography has an opposite sign of the sGVD with respect to the elliptic region. In the conservative long-wavelength limit, this corresponds to the hyperbolic NLSE that does not have wave collapses~\cite{Rasmussen_1986BlowUp}. Indeed, we observe that the spatio-temporal diagram [Fig.~\ref{fig:pic3}(d)] does not demonstrate any extreme events, showing slow (with respect to the elliptic case) incoherent dynamics. Further, comparing the NDR [Fig.~\ref{fig:pic3}(e)] with the elliptic case, we show how the mode occupancy differs. 
In the vicinity of $\mu=0$, the normal sGVD suppresses the photon transfer along the $k$ axis. Nevertheless, the photon transfer to other supermodes is stimulated in the area where the line crosses the lower parabolas, resulting in the generation of dispersive waves~\cite{tikan2021emergent,komagata2021dissipative}. Reconstructing the supermode NDR (Fig.~\ref{fig:pic3}(f)) for $\mu=25$ comb line [the average crossing position in Fig.~\ref{fig:pic3}(e)], we observe the predominant population of the center of the band.

\section{Conclusion}
We theoretically described nonlinear interactions in lattices of photonic microresonators. Considering a simplified model of equally coupled resonators, we demonstrated that this system possesses a 2D dispersion surface and can be described in the long wavelength limit by the 2D LLE at its local extrema.
Different parts of the dispersion surface correspond to two fundamentally different regimes of operation: elliptic and hyperbolic. 
This corresponds to equal and opposite signs of the dispersion, respectively.
Simulating the full set of coupled LLEs, we have demonstrated nonlinear effects inherent to 2D systems which includes hexagonal pattern formation and wave collapses in the chaotic state. Extending these findings to topological lattices, specifically the Su-Schrieffer-Heeger model, we observe nonlinear edge-to-bulk scattering, revealing the loss of topological protection in the presence of four-wave mixing between bands. In summary, our theory sheds light on nonlinear interactions in integrated
photonic lattices and will be helpful for future investigations of multi-mode systems with complex band structures and different topological properties. 

\begin{acknowledgments}
The authors thank Prof. Turitsyn for fruitful discussions.
This publication was supported by Contract18AC00032 (DRINQS) from the Defense Advanced Research Projects Agency (DARPA), Defense Sciences Office (DSO). This material is based upon work supported by the Air Force Office of Scientific Research under award number FA9550-19-1-0250. This work was further supported by the European Union’s Horizon 2020 Research and Innovation Program under the Marie Skłodowska-Curie grant agreement 812818 (MICROCOMB), and by the Swiss National Science Foundation under grant agreement 192293.
\end{acknowledgments}

\section*{Methods}
\textbf{Numerical simulations}. The nonlinear dynamics in the chain of 20 coupled resonators is modeled using step-adaptative Dormand-Prince Runge-Kutta method that is implemented in Python-based library PyCORE https://github.com/ElKosto/PyCORe/tree/PyCORe++ with included integrator from Numerical Recipes 3. The normalized parameters of the simulated system in Eq.~(\ref{SIeq:normalized_coupled_LLEs}) are: $d_2=0.04$, $j=10.13$. To trigger nonlinear dynamics, we  scan the laser from blue
to red detuned side of the resonance. In general, we discretize the system using $512$ sampling points for azimuthal coordinate $\varphi$ of each resonator, but to resolve the regularized wave collapse, we consider $1024$ sampling points.

\bibliography{apssamp}% Produces the bibliography via BibTeX.
%\begin{thebibliography}{36}%

%\end{thebibliography}%

\end{document}